\documentstyle[preprint,aps]{revtex}

\begin{document}
\draft

\title{Quantized Fields and Chronology Protection}
\author{William A. Hiscock\cite{WAH}}
\address{Department of
Physics, Montana State University, Bozeman, Montana 59717}

\date{September 21, 2000}

\maketitle

\begin{abstract}
Several recent possible counterexamples to the Chronology
Protection Conjecture are critically examined. The ``adapted''
Rindler vacuum state constructed by Li and Gott for a conformal
scalar field in Misner space is extended to nonconformally
coupled and self-interacting scalar fields. For these fields, the
vacuum stress-energy always diverges on the chronology horizons.
The divergence of the vacuum stress-energy on Misner space
chronology horizons cannot be generally avoided by choosing a
Rindler-type vacuum state.

\end{abstract}
\pacs{PACS number(s): 04.62.+v}

Over the last decade, there has been substantial interest in
whether it is possible, within the known body of physical law, to
create closed timelike curves (CTC) in a spacetime which is
initially free of such objects~\cite{mty88,gott91,thorne93}. In
colloquial terms, the question is whether it is possible in
principle to construct a ``time machine''. The main impediment
found to such a construction is the divergence of the vacuum
stress-energy of quantized fields in such
spacetimes~\cite{his82,kim91}. This divergence takes place on the
{\it chronology horizon}, the null surface beyond which CTCs
first form. It is believed (but not proven) that the
gravitational backreaction to such a diverging stress-energy
would alter the spacetime in such a way as to prevent the
formation of CTCs. The generic notion that nature will not allow
the formation of CTCs is embodied in Hawking's Chronology
Protection Conjecture (CPC): {\it The laws of physics do not
allow the appearance of closed timelike curves} \cite{hawk92}.

Recently, a number of examples have been found of spacetimes
containing CTCs in which the vacuum stress-energy tensor of a
particular quantum field does not diverge on the chronology
horizon~\cite{sush97,boul92,tanhis95,cass97,ligott98}. These
cases have been interpreted by some \cite{cass97,ligott98} as
possible counterexamples to the Chronology Protection Conjecture,
at least in the form where the vacuum stress-energy of quantized
matter fields is the agent which protects chronology.

However, to be taken seriously, proposed counterexamples to  the
CPC should have to satisfy the same sort of criteria that are
used to evaluate potential counterexamples to the Cosmic
Censorship Hypothesis \cite{Wald97}. To begin, let us assume that
the divergence of $\langle T_{\mu}^{\nu} \rangle$ on chronology
horizons is in fact the mechanism of chronology protection. Then,
if one finds a combination of quantized field(s), vacuum state,
and spacetime such that $\langle T_{\mu}^{\nu} \rangle$ does not
diverge on the chronology horizon, that combination can only be
considered a valid counterexample to the CPC if:

(1) the non-divergence of $\langle T_{\mu}^{\nu} \rangle$ holds
on an open set of spacetime metrics. Counterexamples must not
depend on ``fine-tuning'' of metric parameters or topological
identification scales.

(2) the vacuum stress-energy does not diverge for a collection of
interacting realistic fields. Counterexamples must not depend on
special field properties (e.g., being conformally invariant,
massless, or free).

Condition (1) is general; condition (2) applies only to the
extent that the divergence of quantized field's vacuum
stress-energy is considered to be the mechanism of chronology
protection. The recent examples in which the vacuum stress-energy
is regular on the chronology horizon violate one or both of these
conditions.

For example, Boulware~\cite{boul92} and Tanaka and Hiscock~
\cite{tanhis95} showed that a massive scalar field will have
regular vacuum stress-energy on the chronology horizon of Grant
space~\cite{grant}, provided the field mass is sufficiently
large. In this case, one might argue that the first condition
above is at least partially satisfied, since no fine-tuning of
the Grant space parameters is required to render the vacuum
stress-energy finite. However, the second condition is obviously
violated, since not all quantized fields in the real world are
massive.

As a second example, Sushkov~\cite{sush97} demonstrated that a
complex automorphic massless scalar field would have a
nondivergent vacuum stress-energy on the chronology horizon of
Misner space \cite{Misner} if the automorphic parameter (the
angle by which the complex field is rotated upon topological
identification) has a special value. This violates the first
condition above, since the automorphic parameter must be
fine-tuned to eliminate the divergence in the vacuum
stress-energy. In addition, one could not expect all fields in
nature to be free massless automorphic fields; as Sushkov points
out, the addition of any interaction terms will likely restore
the divergence in the vacuum stress-energy.

On the other hand, Cassidy~\cite{cass97} and Li and Gott
~\cite{ligott98,li99} have shown that there exists a quantum
state in Misner space, an ``adapted'' Rindler vacuum state, for
which the vacuum stress-energy of conformally invariant fields is
finite, in fact precisely zero, provided the Misner space
identification scale is chosen to have a unique special value.
They have indicated that they believe this may serve as a
counterexample to the CPC, or at least to the idea that the
divergence of the vacuum energy of quantized fields can protect
chronology.

This ``counterexample'' violates the first condition above, since
the vacuum stress-energy is nondivergent only for a single value
of the Misner identification scale, a set of measure zero.

In this Letter, I demonstrate that the adapted Rindler vacuum
also violates the second condition. I show that the (Rindler)
vacuum stress-energy of a nonconformally coupled scalar field, or
a conformally coupled massless field with a $\lambda \phi^4$
self-interaction will diverge on the chronology horizon for all
values of the Misner identification scale. In addition, the
vacuum polarization of the field, $\langle \phi^2 \rangle$,
diverges in all cases, even for the conformally invariant case
examined by Li and Gott. Hence, the regular behaviour found by
Cassidy and Li and Gott holds only for a conformally invariant,
non-interacting field, and only for the stress-energy tensor.
While some fields in nature (e.g., the electromagnetic field,
before interactions are added) are conformally invariant, others
-- notably gravity itself -- are not; and interactions are the
rule, not the exception. All calculations are performed in the
Lorentzian signature spacetime, avoiding any conceivable
ambiguity associated with regularization in the Euclidean sector
~\cite{li99}.

Misner space is constructed from Minkowski space by identifying
the points:
\begin{equation}
     (t,x,y,z) \leftrightarrow (t\cosh nb+x\sinh nb, x\cosh nb+t\sinh nb,
      y, z)
\label{ident}
\end{equation}
where $(t,x,y,z)$ are the usual Minkowski Cartesian coordinates, $b$
is an arbitrary positive constant, and $n$ is an integer.
The identifications take on a
simpler form in Misner (equivalently, flat Kasner, or Rindler) coordinates,
$(\eta,\zeta,y,z)$, where
\begin{equation}
 t=\zeta\cosh\eta,~~~x=\zeta\sinh\eta~~~.
\label{xform}
\end{equation}
The metric in these coordinates takes the form
\begin{equation}
ds^2\,=\,-\,d\zeta^2\,+\,\zeta^2\,d\eta^2\,+\,dy^2\,+\,dz^2~~~,
\label{misner1}
\end{equation}
and the points which are identified are now simply
\begin{equation}
     (\zeta,\eta,y,z)\,\,\leftrightarrow\,\,(\zeta,\eta+nb,y,z)~~~.
\label{misident}
\end{equation}.
The Misner space coordinates cover only the past ($P$, with $t >
|x|$) and future ($F$, with $t < |x|$) quadrants of Minkowski
space. There are two null surfaces $\zeta=0$ (corresponding to
$t=x$ and $t=-x$) which are Cauchy and chronology horizons. The
metric may be analytically extended across the boundaries at
$\zeta=0$; the maximal extension of Misner space is obtained by
performing both extensions, although at the cost of obtaining a
non-Hausdorff spacetime with a quasiregular singularity at the
origin $ t=x=0$ \cite {ES}. Due to the topological
identifications of Eq.(\ref{ident}), the extended spacetime now
contains regions of closed timelike curves, namely the right
($R$, $x > |t|$) and left ($L$, $x < |t|$) quadrants of Minkowski
space. In these regions, the $\eta$ and $\zeta$ coordinates
reverse roles, with $\zeta$ becoming a timelike coordinate and
$\eta$ spacelike,
\begin{equation}
ds^2\,=\,-\,\zeta^2d\eta^2\,+\,\,d\zeta^2\,+\,dy^2\,+\,dz^2~~~,
\label{misner2}
\end{equation}
with the relation to Minkowski coordinates now being $t=\zeta
\sinh \eta, x = \zeta \cosh \eta$ in $R$ and $L$.

Consider now the vacuum stress-energy of a non-conformally
coupled quantized massless scalar field in Misner space. The
vacuum stress-energy tensor may be written in terms of the
renormalized Hadamard function as:
\begin{eqnarray}
   \langle T_{\mu\nu}\rangle_{\rm ren}={1\over2}\lim_{X^\prime\rightarrow
   X}\left[(1-2\xi)\nabla_\mu\nabla_{\nu^\prime}
   +(2\xi - {1 \over 2})g_{\mu\nu}\nabla_\alpha\nabla^{\alpha^\prime}
   -2\xi \nabla_\mu\nabla_\nu \right] G^{(1)}_{\rm ren}~~~,
\label{tmndef}
\end{eqnarray}
where $G^{(1)}_{\rm ren}$ is the renormalized Hadamard function. For the
adapted Rindler vacuum state considered by Li and Gott, the renormalized
Hadamard function is obtained by taking the sum over Misner identifications
of the image sources for the Rindler Hadamard function \cite{dow78},
\begin{equation}
   G^{(1)}(X,X^\prime)= {1\over2\pi^2}\sum_{n=-\infty}^{\infty}
   {\gamma\over\zeta\zeta^\prime
   \sinh\gamma[-(\eta-\eta^\prime+nb)^2+\gamma^2]} \; \; ,
\label{GRind}
\end{equation}
where
\begin{equation}
    \cosh\gamma = {\zeta^2 + \zeta^{\prime 2} +
    (y-y^\prime)^2+(z-z^\prime)^2 \over 2 \zeta \zeta^\prime} \;
    \; ,
\label{gamma}
\end{equation}
and subtracting the Minkowski vacuum Hadamard function,
\begin{equation}
   G_{\rm ren}^{(1)}=G^{(1)}-G_{\rm Mink}^{(1)}~~~,
\label{Gren}
\end{equation}
where, as usual,
\begin{equation}
   G_{\rm Mink}^{(1)}(X,X^\prime)={1\over2\pi^2}\;\;
   {1\over
   [-(t-t^\prime)^2+(x-x^\prime)^2+(y-y^\prime)^2+(z-z^\prime)^2]}~~.
\label{GMink}
\end{equation}
Note that the Hadamard functions are independent of the curvature
coupling of the scalar field, $\xi$. This is because the mode
equation for the field is independent of $\xi$ in any spacetime
in which the Ricci curvature scalar vanishes, in particular, in
flat space. The stress-energy tensor, defined via
Eq.(\ref{tmndef}), however, is {\it not} independent of $\xi$,
even in flat space.

It is a simple matter to take the required derivatives and coincidence limit
of $G^{(1)}_{\rm ren}$ to obtain the components of $\langle T_{\mu\nu}
\rangle$; the result is
\begin{equation}
   \langle T_{\zeta\zeta}\rangle={1 \over {3\zeta^2}} \langle T_{\eta\eta}
   \rangle ={{(b^2+4\pi^2)[4\pi^2-b^2-10(1-6\xi)b^2]} \over
   {1440\pi^2 b^4 \zeta^4}}~~~,
\label{Tmn1}
\end{equation}
\begin{equation}
   \langle T_{yy}\rangle= \langle T_{zz}\rangle =
   {{(b^2+4\pi^2)[4\pi^2-b^2+20(1-6\xi)b^2]} \over
   {1440\pi^2 b^4 \zeta^4}}~.
\label{Tmn2}
\end{equation}
It is now easy to see that there is no value of $b$ for which all the
components of $\langle T_{\mu\nu}\rangle$ will be regular on the
chronology horizon at $\eta = 0$, unless the
field is conformally coupled, $\xi=1/6$. From Eq.(\ref{Tmn1}),
$T_{\eta\eta}$ and $T_{\zeta\zeta}$ will be regular at the chronology
horizon only if
\begin{equation}
     b^2 = {4 \pi^2 \over 11 - 60 \xi}~~,
\label{cond1}
\end{equation}
while $T_{yy}$ and $T_{zz}$ will be regular on the horizon only if
\begin{equation}
     b^2 = {4 \pi^2 \over 120\xi-19}~~.
\label{cond2}
\end{equation}
Equations (\ref{cond1}) and (\ref{cond2}) can only be
simultaneously satisfied if $ b = 2\pi, \xi = 1/6$. Thus, the
stress-energy of any nonconformally coupled ($\xi \neq 1/6$)
scalar field will diverge on the chronology horizon in the usual
$\zeta^{-4}$ manner, regardless of the particular value of the
identification scale $b$.

Next consider a scalar field with a $\lambda \phi^4$
self-interaction, again in Misner space in the adapted Rindler
vacuum state. For simplicity, only the case where the free
field's vacuum stress-energy is nondivergent will be treated,
i.e., a massless, conformally coupled ($\xi = 1/6$)
self-interacting field. To first order in the self-coupling
constant $\lambda$, the vacuum stress-energy may be written as
\begin{equation}
\langle T_{\mu \nu} \rangle = \langle T_{\mu \nu}^{\rm free}
\rangle + \langle T_{\mu \nu}^{\rm self-int} \rangle \; \; ,
\label{tmnint1}
\end{equation}
where $\langle T_{\mu \nu}^{\rm free} \rangle$ is the vacuum
stress-energy of the free field. In Misner space, the
self-interaction stress-energy is uniquely determined by the
self-interaction energy density together with the requirements
that the stress-energy tensor be conserved ($\langle
T_\mu^\nu\rangle_{; \, \nu} = 0$), traceless ($\langle T_\mu^\mu
\rangle = 0$, which follows as the field is conformally invariant
and the spacetime is flat), and symmetric in the $y-z$ plane
($\langle T_{yy} \rangle = \langle T_{zz} \rangle$). Working in
the $R$ quadrant, these conditions yield:
\begin{eqnarray}
    \langle T_{\eta \eta}^{\rm self-int} \rangle &=& \zeta^2 \rho \;
    \; , \nonumber \\
    \langle T_{\zeta \zeta}^{\rm self-int} \rangle
    &=& - {1 \over \zeta}  \int \rho \; d\zeta  \; \; , \nonumber \\
    \langle T_{yy}^{\rm self-int} \rangle & = & \langle T_{zz}^{\rm
    self-int} \rangle = {1 \over 2} \left ( \rho + {1 \over \zeta}
    \int \rho \; d\zeta \right) \;\; ,
\label{selfintform}
\end{eqnarray}
where $\rho$ is the self-interaction energy density. The value of
$\rho$ may be determined using the methods of Ford \cite{Ford}
and Kay \cite{Kay},
\begin{equation}
\rho =  { \lambda \over 4!} \langle \phi^4 \rangle =  {\lambda
\over 8} \langle \phi^2 \rangle^2 \; \; . \label{rho}
\end{equation}
In principle, before evaluating $\langle T_{\mu \nu}^{\rm
self-int} \rangle$, the field mass $M$, curvature coupling $\xi$,
self-coupling $\lambda$, and the wavefunction must be
renormalized. However, the mass counterterm at first order in
$\lambda$ is zero for a massless field, the curvature coupling
counterterm vanishes for conformal coupling at first order, and
both self-coupling and wavefunction renormalization are second
order in $\lambda$ and hence can be ignored here. We can thus
proceed to evaluate $\langle \phi^2 \rangle$ using the Hadamard
function of Eq.(\ref{Gren}),
\begin{equation}
\langle \phi^2 \rangle = {1 \over 2} \lim_{X' \rightarrow
X^\prime} G_{\rm ren}^{(1)} (X,X^\prime) \; \; .
\label{vpdef}
\end{equation}
The actual evaluation of $\langle \phi^2 \rangle$, often called
the ``vacuum polarization'' of a scalar field, is then a simple
exercise, which yields
\begin{equation}
\langle \phi^2 \rangle = { - 1 \over 48 \pi^2 \zeta^2} \left [ 1
+ \left ( {2 \pi \over b} \right )^2 \right ] \; \; ,
\label{phi2}
\end{equation}
which clearly diverges at the chronology horizons at $\eta =0$
for all values of the Misner identification parameter $b$. Thus,
the vacuum polarization, $\langle \phi^2 \rangle$, is always
divergent on the chronology horizons, even for the free field.

The self-interaction contribution to vacuum stress-energy for a
self-interacting, massless, conformally coupled scalar field in
the adapted Rindler vacuum to first order in $\lambda$ is then,
by Eqs.(\ref{selfintform}), (\ref{rho}) and (\ref{phi2}),
\begin{equation}
 \langle T_{\mu \nu}^{\rm self-int} \rangle = {\lambda \over
 {18432 \pi^4 \zeta^4}} \left[ 1 + \left( {2 \pi \over b} \right)^2
 \right]^2 {\rm diag}\left(\zeta^2, {1 \over 3}, {1 \over 3}, {1 \over 3}
 \right)\; \; ,
\label{tmnselfint}
\end{equation}
where the coordinates are ordered $(\eta,\zeta,y,z)$.

The total vacuum stress-energy is this term plus the free-field
vacuum stress-energy for the conformal massless field, which is
given by Eqs.(\ref{Tmn1}-\ref{Tmn2}) with $ \xi = 1/6 $. In the
special case examined by Li and Gott, with the Misner
identification scale set to $b = 2 \pi$,  the free field vacuum
stress-energy vanishes everywhere. However, the self-interaction
component of the vacuum stress-energy, as calculated here to
first order in $\lambda$, diverges on the chronology horizon for
all values of $b$. The divergence has the identical $\zeta^{-4}$
dependence that is found generally. Since this calculation is
done within the context of first-order perturbation theory, it is
conceivable that this divergence would vanish in the full
(nonperturbative) theory. It seems more likely, however, that
miraculous cancellations do not occur, and thus that the
divergence would persist in the full nonperturbative theory.

In conclusion, the few examples that have been found of a
combination of quantized field, vacuum state, and spacetime that
yield a non-divergent $\langle T_\mu^\nu \rangle$ on the
chronology horizon all involve special choices of fields and
spacetime properties. They are not robust when examined in a
larger context. In particular, the adapted Rindler vacuum in
Misner space, shown to have vanishing vacuum stress-energy for
massless conformally coupled scalar field by Li and Gott, has
here been shown here to have divergent stress-energy on the
chronology  horizons for nonconformally coupled fields and also
for self-interacting fields. The adapted Rindler vacuum in Misner
space does not seem to provide a convincing counterexample to the
Chronology Protection Conjecture.

\acknowledgments I wish to thank Paul Anderson for helpful
discussions. This research was supported by NSF grant
PHY-9734834.


\begin{references}
\bibitem[*]{WAH}electronic mail address: hiscock@physics.montana.edu
\bibitem{mty88} M. S. Morris, K. S. Thorne, and U. Yurtsever,
    Phys. Rev. Lett. {\bf 61}, 1446 (1988).
\bibitem{gott91} J. R. Gott, Phys. Rev. Lett. {\bf 66}, 1126 (1991).
\bibitem{thorne93} K. S. Thorne, in {\sl General Relativity and
    Gravitation 1992: Proceedings of the 13th International
    Conference on General Relativity and Gravitation}, edited by R.
    J. Gleiser, C. N. Kozameh and O. M. Moreschi (IOP Publishing,
    Bristol, 1993), p. 295.
\bibitem{his82} W. A. Hiscock and D. A. Konkowski, Phys. Rev. D
    {\bf 26}, 1225 (1982).
\bibitem{kim91} S. -W. Kim and K. S. Thorne, Phys. Rev. D {\bf 43},
    3929 (1991).
\bibitem{hawk92} S. W. Hawking, Phys. Rev. D {\bf 46}, 603 (1992).
\bibitem{sush97} S. V. Sushkov, Class. Quantum Grav. {\bf 14}, 523 (1997).
\bibitem{boul92} D. G. Boulware, Phys. Rev. D {\bf 46}, 4421
    (1992).
\bibitem{tanhis95} T. Tanaka and W. A. Hiscock, Phys. Rev. D {\bf
    52}, 4503 (1995).
\bibitem{cass97} M. J. Cassidy, Class. Quantum Grav. {\bf 14},
    3031 (1997).
\bibitem{ligott98} L.-X. Li and J. R. Gott, Phys. Rev. Lett. {\bf
    80}, 2980 (1998).
\bibitem{Wald97} R.\ M.\ Wald, unpublished (gr-qc/9710068).
\bibitem{grant} J. D. E. Grant, Phys. Rev. D {\bf 47}, 2388 (1993).
\bibitem{Misner} C. W. Misner, in {\sl Relativity Theory
    and Astrophysics I: Relativity and Cosmology}, edited by J.\
    Ehlers, Lectures in Applied Mathematics, Vol. 8 (American
    Mathematical Society, Providence, 1967), p. 160.
\bibitem{li99} L.-X. Li, Phys. Rev. D {\bf 59 }, 084016 (1999).
\bibitem{ES} G.\ F.\ R.\ Ellis and B.\ G.\ Schmidt, Gen.\
    Relativ.\ Gravit.\ {\bf 8}, 915 (1974).
\bibitem{dow78} J. S. Dowker, Phys. Rev. D {\bf 18}, 1856 (1978).
\bibitem{Tanaka} The self-interacting field in Misner space
has been analyzed for the Minkowski vacuum state by T. Tanaka,
Montana State University Ph.D. dissertation, 1997 (unpublished).
\bibitem{Ford} L.\ H.\ Ford, Proc. R. Soc. (London) A {\bf 368},
305 (1979).
\bibitem{Kay} B.\ S.\ Kay, Phys. Rev. D {\bf 20}, 3052 (1979).
\end{references}
\end{document}